\documentclass[english,preprint,aps,pra,superscriptaddress]{revtex4-1}
\usepackage{lmodern}
\usepackage[T1]{fontenc}
\usepackage[latin9]{inputenc}
\usepackage{babel}
\usepackage{amsmath}
\usepackage{amssymb}
\usepackage{graphicx}
\usepackage[unicode=true,pdfusetitle,
 bookmarks=true,bookmarksnumbered=false,bookmarksopen=false,
 breaklinks=true,pdfborder={0 0 1},backref=false,colorlinks=false]
 {hyperref}
\usepackage{breakurl}

\makeatletter

\newcommand{\lyxdot}{.}

\makeatother

\begin{document}

\title{Understanding Detailed Balance for an Electron\textendash Radiation
System Through Mixed Quantum\textendash Classical Electrodynamics}

\author{Hsing-Ta Chen}
\email{hsingc@sas.upenn.edu}

\selectlanguage{english}%

\affiliation{Department of Chemistry, University of Pennsylvania, Philadelphia,
Pennsylvania 19104, U.S.A.}

\author{Tao E. Li}

\affiliation{Department of Chemistry, University of Pennsylvania, Philadelphia,
Pennsylvania 19104, U.S.A.}

\author{Abraham Nitzan}

\affiliation{Department of Chemistry, University of Pennsylvania, Philadelphia,
Pennsylvania 19104, U.S.A.}

\author{Joseph E. Subotnik}

\affiliation{Department of Chemistry, University of Pennsylvania, Philadelphia,
Pennsylvania 19104, U.S.A.}
\begin{abstract}
We investigate detailed balance for a quantum system interacting with
thermal radiation within mixed quantum\textendash classical theory.
For a two-level system coupled to classical radiation fields, three
semiclassical methods are benchmarked: (1) Ehrenfest dynamics over-estimate
the excited state population at equilibrium due to the failure of
capturing vacuum fluctuations. (2) The coupled Maxwell\textendash Bloch
equations, which supplement Ehrenfest dynamics by damping at the full
golden rule rate, under-estimate the excited state population due
to double-counting of the self-interaction effect. (3) Ehrenfest+R
dynamics recover detailed balance and the correct thermal equilibrium
by enforcing the correct balance between the optical excitation and
spontaneous emission of the quantum system. These results highlight
the fact that, when properly designed, mixed quantum\textendash classical
electrodynamics can simulate thermal equilibrium in the field of nanoplasmonics. 
\end{abstract}
\maketitle

\section{Introduction\label{sec:Introduction}}

For a quantum system in contact with a thermal bath at temperature
$T$, detailed balance implies that the equilibrium population of
quantum state $i$ with energy $E_{i}$ should satisfy the Boltzmann
distribution: $P_{i}\propto e^{-\beta E_{i}}$ where $\beta=1/k_{B}T$
and $k_{B}$ is the Boltzmann constant. This principle of detailed
balance is one of the most fundamental aspects underlying chemical
kinetics and absorption/emission of radiation\citep{onsager_reciprocal_1931,loudon_quantum_2000,milonni_quantum_2013}.
For example, the Onsager reciprocal relations in thermodynamics and
Einstein's famous A and B coefficients in his theory of radiation
are both based on detailed balance. Thus, maintaining detailed balance
is considered an important measure of validity for any theoretical
model of coupled system\textendash bath equilibrium. For decades,
achieving the correct equilibrium populations has been a long-standing
challenge for semiclassical simulations of electronic\textendash nuclear
dynamics\citep{parandekar_mixed_2005,parandekar_detailed_2006,schmidt_mixed_2008,bastida_ehrenfest_2007}
and electron\textendash radiation interactions\citep{milonni_semiclassical_1976,slavcheva_coupled_2002,ruggenthaler_quantum-electrodynamical_2014}.

Within the field of electronic\textendash nuclear dynamics, semiclassical
simulations rely on a mixed quantum\textendash classical framework
that treats the electronic/molecular system with quantum mechanics
and the bath degrees of freedom with classical mechanics. Within this
framework, recovering detailed balance requires both energy conservation
of the entire system \emph{and} the correct energy exchange rates
between the quantum and classical subsystems. Already a decade ago,
Tully showed that the equilibrium populations of a coupled electron\textendash nuclei
problem as attained by Ehrenfest dynamics deviates from the correct
Boltzmann distribution when the nuclear bath temperature decreases\citep{parandekar_mixed_2005}.
This deviation can be attributed to the deficiency of Ehrenfest mean-field
theory to properly account for non-adiabatic electronic transitions
(even though the total energy is conserved)\citep{parandekar_detailed_2006}.
To capture non-adiabatic effects within a mixed quantum\textendash classical
framework, the most common solution is either to design a stochastic
mechanism to simulate electronic transitions, such as surface hopping
algorithms\citep{tully_molecular_1990,tully_mixed_1998,schmidt_mixed_2008,jain_surface_2015-1,bellonzi_assessment_2016,sifain_communication:_2016},
or to introduce binning as in the symmetrical quasi-classical (SQC)
approach\citep{cotton_symmetrical_2013,miller_communication:_2015},
both of which can almost recover detailed balance for coupled electron\textendash nuclei
equilibrium. 

If we now turn to electron\textendash radiation dynamics, the equations
of motion within a mixed quantum\textendash classical framework are
formally similar to those for electronic\textendash nuclear dynamics.
And yet, because thermal radiation fields cannot be properly modeled
by classical electrodynamics (due to the notorious blackbody radiation
problem), the applicability of Tully's argument to electron\textendash radiation
equilibrium is unclear. And more generally, the feasibility of semiclassical
techniques to model electrodynamics and to reach detailed balance
quantitatively remains an open question. While many semiclassical
schemes for electrodynamics have been introduced\citep{milonni_semiclassical_1976,miller_classical/semiclassical_1978},
including the coupled Maxwell\textendash Bloch equations\citep{castin_maxwell-bloch_1995,ziolkowski_ultrafast_1995,slavcheva_coupled_2002},
mean-field Ehrenfest dynamics\citep{li_mixed_2018,li_comparison_2019,hoffmann_light-matter_2018,hoffmann_capturing_2019},
Ehrenfest+R method\citep{chen_ehrenfest+r_2019,chen_ehrenfest+r_2019-1,chen_predictive_2019-1},
and quantum electrodynamical density functional theory (QEDFT)\citep{ruggenthaler_quantum-electrodynamical_2014,flick_atoms_2017,ruggenthaler_quantum-electrodynamical_2018},
the capacity of these semiclassical approaches to recover detailed
balance has never been fully benchmarked. It is a common presumption
that mixed quantum\textendash classical electrodynamics cannot satisfy
detailed balance due to the failure of classical electrodynamics to
describe the blackbody radiation spectrum\textemdash modeling thermal
radiation fields classically should inevitably lead to the incorrect
Rayleigh\textendash Jeans spectrum\citep{boyer_derivation_1969,milonni_semiclassical_1976,milonni_quantum_2013}.

Nevertheless, Boyer recently showed that detailed balance can be achieved
by considering a fully classical model composed of a classical charged
harmonic oscillator coupled to a set of classical electromagnetic
(EM) fields\citep{boyer_equilibrium_2018}. Boyer's analysis is based
on relativistic classical mechanics and random electrodynamics theory
that characterizes the fluctuations of thermal radiation using an
ensemble of classical EM fields with a random phase\citep{boyer_derivation_1969,boyer_random_1975}.
As shown by Boyer, this classical model of thermal radiation fields
can overcome the failure of classical electrodynamics and recover
the correct Planck spectrum and zero-point energy\citep{boyer_derivation_1969,boyer_equilibrium_1976,boyer_random_1975,boyer_understanding_2016}
. As such, there is at least one example for how classical electrodynamics
can recover the quantum blackbody radiation without invoking the quantization
of light as photons. 

In this paper, our goal is to employ Boyer's framework for classically
modeling thermal radiation fields and then evaluate the capacity of
various mixed quantum\textendash classical methods to recover detailed
balance for the electron\textendash radiation equilibrium; note that,
unlike Boyer, we will not invoke relativity but rather treat the electronic
subsystem with quantum mechanics. Our approach will be to model an
electronic two-level system (TLS) coupled to a bath of incoming radiation
EM fields at temperature $T$ and then to compare the equilibrium
population with the Boltzmann distribution. This paper is organized
as follows. In Sec.~\ref{sec2}, we set up the model Hamiltonian
and formulate the boundary conditions for thermal radiation fields.
In Sec.~\ref{sec3}, we briefly review three semiclassical approaches
for electron\textendash radiation dynamics. In Sec.~\ref{sec4},
we compare the mixed quantum\textendash classical equilibrium as attained
by these three different semiclassical models against the correct
Boltzmann distribution. We conclude with an outlook for the future
in Sec.~\ref{sec5}.

For notation, we use a bold symbol to denote a space vector $\mathbf{r}=x\boldsymbol{\epsilon}_{x}+y\boldsymbol{\epsilon}_{y}+z\boldsymbol{\epsilon}_{z}$
where $\boldsymbol{\epsilon}$ denotes a unit vector in Cartesian
coordinate. We use $\int dv=\int dxdydz$ for integration over 3D
space. We work below in SI units.

\section{The Hamiltonian and boundary conditions\label{sec2}}

For a model of electron\textendash radiation equilibrium, we consider
an electronic TLS coupled to thermal radiation fields in a 3D space.
The TLS Hamiltonian is $\widehat{H}_{s}=\hbar\omega_{0}\left|e\right\rangle \left\langle e\right|$
where the ground state $\left|g\right\rangle $ and the excited state
$\left|e\right\rangle $ are separated by an energy difference $\hbar\omega_{0}$.
For a quantum electrodynamics (QED) description, the TLS is coupled
to a set of photon fields that describe thermal radiation fields.
Such a model should reach thermal equilibrium when the optical excitation
by thermal radiation fields are balanced against the spontaneous emission
of the TLS. In the end, the equilibrium populations should follow
the Boltzmann distribution (obeying the principle of detailed balance)
\begin{equation}
P_{e}/P_{g}=e^{-\beta\hbar\omega_{0}}.
\end{equation}

In general, recovering detailed balance requires an accurate treatment
of spontaneous emission. It is well-known that, in QED, spontaneous
emission includes two subprocesses: self-interaction and the vacuum
fluctuations\citep{dalibard_vacuum_1982,dalibard_dynamics_1984}.
Self-interaction is the subprocess induced by the emitted field interacting
back with the TLS. The vacuum fluctuations are purely quantum effects
that arise from the zero-point energy of the quantum EM fields inducing
decay of the TLS.

\subsection{Semiclassical Electronic Hamiltonian}

For mixed quantum\textendash classical electrodynamics, the TLS is
described by an electronic reduced density matrix $\rho\left(t\right)$
while the EM fields, $\mathbf{E}\left(\mathbf{r},t\right)$ and $\mathbf{B}\left(\mathbf{r},t\right)$,
are classical. The electric dipole coupling is $\widehat{V}\left(t\right)=-\int\mathrm{d}v\widehat{\mathbf{P}}\left(\mathbf{r}\right)\cdot\mathbf{E}\left(\mathbf{r},t\right)$,
which couple the TLS to the total electric field. The polarization
density operator of the TLS is defined as $\widehat{\mathbf{P}}\left(\mathbf{r}\right)=\mathbf{d}\left(\mathbf{r}\right)\left(\left|e\right\rangle \left\langle g\right|+\left|g\right\rangle \left\langle e\right|\right)$.
We assume that the polarization density takes the form $\mathbf{d}\left(\mathbf{r}\right)=\mu\frac{2e^{-r^{2}/2\sigma^{2}}}{\left(2\pi\sigma^{2}\right)^{3/2}}\boldsymbol{\epsilon}_{z}$
where $\left|\mathbf{r}\right|=r$. Here, $\mu$ is the magnitude
of the transition dipole moment and $\sigma$ represents the length
scale of the TLS. Note that, for simplicity, we set the polarization
to be along the $z$ direction. In practice, we usually make the
long-wavelength approximation, assuming that the wavelength of the
dominant absorption/emission mode ($2\pi c/\omega_{0}$) is much larger
than the size of the TLS (usually on the order of a Bohr radius).
With this assumption, we can approximate the polarization density
as a point dipole, i.e. $\mathbf{d}\left(\mathbf{r}\right)\rightarrow\mu\delta^{3}\left(r\right)\boldsymbol{\epsilon}_{z}$.
For the TLS in vacuum, the spontaneous emission rate is given by
the Fermi's golden rule (FGR) rate, $\kappa=\mu^{2}\omega_{0}^{3}/3\pi\hbar\epsilon_{0}c^{3}$.

Explicitly, the semiclassical electronic Hamiltonian in matrix form
reads
\begin{equation}
\widehat{H}^{\mathrm{el}}=\widehat{H}_{s}+\widehat{V}\left(t\right)=\left(\begin{array}{cc}
0 & v\left(t\right)\\
v\left(t\right) & \hbar\omega_{0}
\end{array}\right),\label{eq:elec_Ham}
\end{equation}
where the electric dipole coupling is $v\left(t\right)=-\mu\boldsymbol{\epsilon}_{z}\cdot\mathbf{E}\left(0,t\right)$.
Here, we can decompose the total electric field as $\mathbf{E}=\mathbf{E}_{\mathrm{in}}+\mathbf{E}_{\mathrm{scatt}}$
in terms of the incoming thermal radiation fields ($\mathbf{E}_{\mathrm{in}}$)
and the scattered field ($\mathbf{E}_{\mathrm{scatt}}$) generated
by the stimulated and spontaneous emission processes. In the end,
the electric dipole coupling includes two terms: (i) the self-interaction
term $\mu\boldsymbol{\epsilon}_{z}\cdot\mathbf{E}_{\mathrm{scatt}}\left(0,t\right)$
and (ii) the optical excitation term $\mu\boldsymbol{\epsilon}_{z}\cdot\mathbf{E}_{\mathrm{in}}\left(0,t\right)$.
As is common, we will assume that the effects of these two processes
are essentially independent\citep{cohen-tannoudji_atom-photon_1998}.

\subsection{Boundary Conditions: Thermal Radiation as Random Electromagnetic
Fields }

The optical excitation term $\mu\boldsymbol{\epsilon}_{z}\cdot\mathbf{E}_{\mathrm{in}}\left(0,t\right)$
represents the coupling between the TLS and the incoming classical
EM fields. As discussed in the introduction, we will employ Boyer's
random electrodynamical model to simulate thermal radiation fields. 

As it was originally formulated\citep{boyer_derivation_1969}, random
electrodynamics constitute a classical model of thermal radiation
fields in terms of a sum over transverse plane waves with a random
phase in 3D space (see Eq.~\eqref{eq:Boyer_Ein}). These transverse
plane waves follow homogeneous Maxwell equations: $\frac{\partial}{\partial t}\mathbf{B}_{\mathrm{in}}=-\boldsymbol{\nabla}\times\mathbf{E}_{\mathrm{in}}$,
$\frac{\partial}{\partial t}\mathbf{E}_{\mathrm{in}}=c^{2}\boldsymbol{\nabla}\times\mathbf{B}_{\mathrm{in}}$.
With this random electrodynamical model of thermal radiation fields,
the optical excitation term for the electric dipole coupling can be
written as a sum over harmonic oscillators with discrete frequency
modes (see Appendix~\ref{appendix_A})

\begin{equation}
\mu\boldsymbol{\epsilon}_{z}\cdot\mathbf{E}_{\mathrm{in}}\left(0,t\right)=\sum_{\omega}\mu A\left(\omega,\beta\right)\cos\left[\omega t+\theta_{\omega}\right].
\end{equation}
The random phase $\theta_{\omega}\in\left[0,2\pi\right)$ is chosen
independently for each frequency $\omega$. Here, the coupling strength
for frequency mode $\omega$ is determined by the spectrum
\begin{equation}
\left|A\left(\omega,\beta\right)\right|^{2}=\frac{2\hbar}{3\pi^{2}\epsilon_{0}c^{3}}\frac{\omega^{3}}{e^{\beta\hbar\omega}-1}d\omega,\label{eq:PlanckSpectrum}
\end{equation}
which corresponds to the true QM energy density within the frequency
range $\left(\omega,\omega+d\omega\right)$, though ignoring zero-point
energy. 

Finally, we emphasize that random electrodynamics theory exploits
an ensemble of classical EM fields to simulate the statistical variations
of a quantum electrodynamical field. We denote $\left\langle \cdot\right\rangle _{\theta}$
as an average with respect to the random phase $\{\theta_{\omega}\}$.
For example, the radiation energy density can be evaluated by $\frac{\epsilon_{0}}{2}\langle\mathbf{E}_{\mathrm{in}}^{2}\rangle_{\theta}+\frac{1}{2\mu_{0}}\langle\mathbf{B}_{\mathrm{in}}^{2}\rangle_{\theta}=\sum_{\omega}U\left(\omega,\beta\right)/V$
and the total electric field is $\langle\mathbf{E}_{\mathrm{in}}\rangle_{\theta}=0$
due to the random phase cancellation. In what follows, we will run
multiple trajectories with $\{\theta_{\omega}\}$ chosen randomly
and average over those trajectories to evaluate physical observables,
such as the average density matrix $\langle\rho\rangle_{\theta}$.

\section{Mixed Quantum\textendash Classical Electrodynamics\label{sec3}}

In this section, we briefly review three semiclassical methods for
simulating electron\textendash radiation dynamics: (1) Ehrenfest dynamics,
(2) the coupled Maxwell\textendash Bloch equations, and (3) the Ehrenfest+R
approach. While these methods treat the optical excitation by the
incoming thermal radiation fields in the same manner, the incorporation
of spontaneous emission effects for the TLS is considered at different
levels of accuracy\citep{li_comparison_2019}. Since we are interested
in the equilibrium population, we focus on the equation of motion
for the TLS below.

\subsection{Ehrenfest dynamics}

Ehrenfest dynamics is the most straightforward, mean\textendash field
semiclassical ansatz for electrodynamics\citep{li_mixed_2018}. The
density matrix of the TLS evolves according to the Liouville equation:
$\dot{\rho}=-\frac{i}{\hbar}[\widehat{H}^{\mathrm{el}},\rho]$. The
scattered EM fields follow the inhomogeneous Maxwell's equations:
$\frac{\partial}{\partial t}\mathbf{B}_{\mathrm{scatt}}=-\boldsymbol{\nabla}\times\mathbf{E}_{\mathrm{scatt}}$,
$\frac{\partial}{\partial t}\mathbf{E}_{\mathrm{scatt}}=c^{2}\boldsymbol{\nabla}\times\mathbf{B}_{\mathrm{scatt}}-\mathbf{J}/\epsilon_{0}$.
Here, the current source $\mathbf{J}=\frac{\partial}{\partial t}\text{Tr}\{\rho\widehat{\mathbf{P}}\}$
is generated by the average polarization of the electronic system.
Within Ehrenfest dynamics, the total energy of the TLS and the scatted
EM fields is conserved. However, as shown in Refs.~\onlinecite{li_mixed_2018,chen_ehrenfest+r_2019},
Ehrenfest dynamics completely ignore the effects of vacuum fluctuations
and fail to capture spontaneous emission quantitatively. More specifically,
for a TLS in vacuum, the population decay rate is $k_{\mathrm{Eh}}(t)=\kappa\rho_{gg}(t)$,
i.e. Ehrenfest dynamics tend to be more accurate in the weak excitation
limit ($\rho_{gg}\rightarrow1$, $\rho_{ee}\rightarrow0$) than in
the strong excitation limit ($\rho_{gg}\rightarrow0$, $\rho_{ee}\rightarrow1$).

\subsection{Coupled Maxwell\textendash Bloch equations}

The most common scheme to improve Ehrenfest dynamics is to add some
phenomenological dissipation to the Liouville equation: $\dot{\rho}=-\frac{i}{\hbar}[\widehat{H}^{\mathrm{el}},\rho]+{\cal L}_{\mathrm{SE}}[\rho]$
(forming the so-called the coupled Maxwell\textendash Bloch equations\citep{castin_maxwell-bloch_1995,ziolkowski_ultrafast_1995})
while the scattered EM fields follow the same equations of motion
as Ehrenfest dynamics. The dissipation term takes the form of a Lindblad
operator: ${\cal L}_{\mathrm{SE}}[\rho]=\kappa\left(\widehat{a}\rho\widehat{a}^{\dagger}-\frac{1}{2}\widehat{a}^{\dagger}\widehat{a}\rho-\frac{1}{2}\rho\widehat{a}^{\dagger}\widehat{a}\right)$
where $\widehat{a}=\left|g\right\rangle \left\langle e\right|$. Note
that the full FGR rate $\kappa$ is used in this phenomenological
dissipation. The Lindblad operator can be derived from a QED calculation
to describe the \emph{overall} effects of spontaneous emission. 

Unfortunately, naively supplementing Ehrenfest dynamics by this phenomenological
damping at the full FGR rate causes many disadvantages for the coupled
Maxwell\textendash Bloch equations. First, the total energy of the
TLS and the radiation fields is not conserved: there is no EM field
emission corresponding to the additional dissipation of the TLS. Second,
the effect of the self-interaction is double-counted: the phenomenological
Lindblad operator with the FGR rate ($\kappa$) ignores the fact that
self-interaction has already been included in Ehrenfest dynamics and
only the vacuum fluctuations must be addressed. As has been shown
recently\citep{li_comparison_2019}, this naive implementation of
the coupled Maxwell\textendash Bloch equations almost always \emph{over-estimates}
the population decay rate of the TLS.

\subsection{Ehrenfest+R approach}

In contract to the coupled Maxwell\textendash Bloch equations, the
Ehrenfest+R approach correctly adds only the effect of vacuum fluctuations
on top of the Ehrenfest dynamics\citep{chen_ehrenfest+r_2019}. Explicitly,
the +R correction enforces three effects: (1)~population relaxation,
which adjusts the electronic population to recover spontaneous emission,
(2)~stochastic dephasing, which introduces a stochastic random phase
to the electronic wavefunction, (3)~EM field rescaling, which enforces
total energy conservation. A detailed implementation of the Ehrenfest+R
method is presented in Ref.~\onlinecite{chen_ehrenfest+r_2019,li_comparison_2019}.

Now, we emphasize that, within Ehrenfest+R dynamics, the scattered
EM fields are described on the same footing as the incoming thermal
radiation within random electrodynamics. Since the stochastic random
phase $\phi$ is introduced by the +R correction, the scattered EM
fields in Ehrenfest+R dynamics constitute an ensemble of classical
EM fields. Thus, in the end, when evaluating physical observables,
we must average over both $\theta$ (from thermal radiation) and $\phi$
(from the +R correction). For example, if we use an electronic wavefunction
to describe the TLS $\left|\psi\left(t\right)\right\rangle =c_{g}\left(t\right)\left|g\right\rangle +c_{e}\left(t\right)\left|e\right\rangle $,
the excited state population of the TLS $P_{e}\left(t\right)=\langle\left|c_{e}\left(t\right)\right|^{2}\rangle_{\phi}\rangle_{\theta}$.

\section{Results and Discussion\label{sec4}}

To reach thermal equilibrium between the TLS and the incoming radiation
fields, we simulate the dynamics of the TLS for long time ($t=200/\kappa$).
We vary the temperature of the thermal radiation ($k_{B}T=1/\beta$)
and compare the excited state populations at equilibrium with the
Boltzmann distribution: $P_{e}=1/(1+e^{\beta\hbar\omega_{0}})$. Here,
we choose $\kappa\ll\omega_{0}$ so as to operate in the FGR regime.
The coupling strength $A\left(\omega,\beta\right)$ is discretized
by $51$ modes with $d\omega=\kappa/300$ near $\omega_{0}$. For
each $\beta$, we run $2000$ trajectories for convergence.

\begin{figure}
\noindent \begin{centering}
\includegraphics{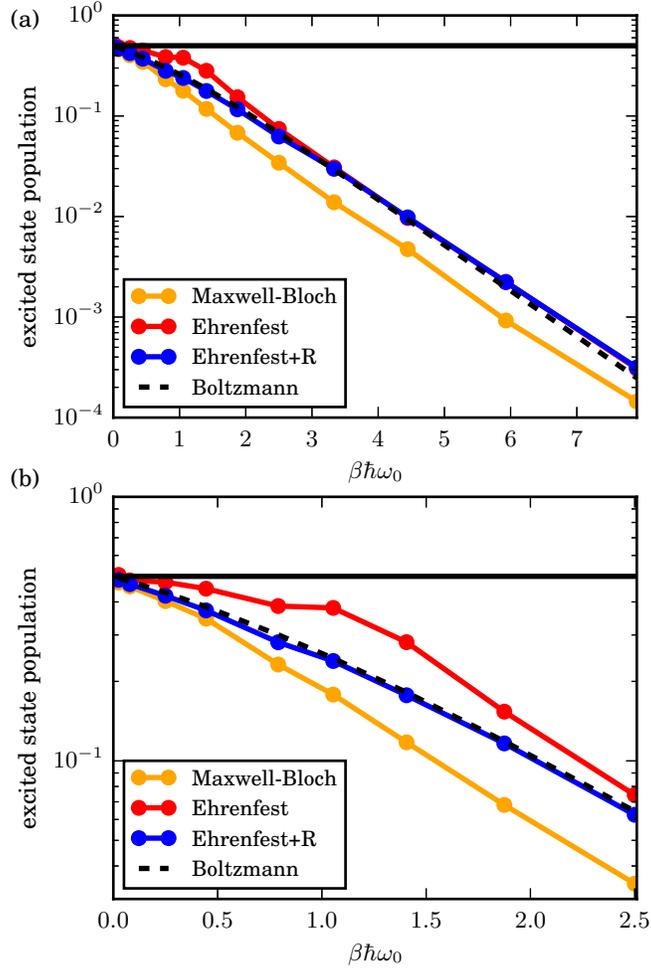}
\par\end{centering}
\caption{Equilibrium populations of the excited state of the TLS as a function
of inverse temperature parameter $\beta\hbar\omega_{0}$. The dashed
black lines are the correct Boltzmann distribution. The solid black
lines correspond to the saturate population at infinite temperature
($P_{e}=P_{g}=0.5$). The solid circles are the average data obtained
by Ehrenfest dynamics (colored red), the coupled Maxwell\textendash Bloch
equations (colored orange), and the Ehrenfest+R approach (colored
blue). (a) For a large range of temperature, Ehrenfest+R dynamics
recover the correct Boltzmann distribution. The coupled Maxwell\textendash Bloch
equations almost always under-estimate equilibrium population due
to the double-counting of the self-interaction. Surprisingly, Ehrenfest
dynamics results agree with the correct Boltzmann distribution at
low temperature. (b) In the high temperature regime, Ehrenfest dynamics
over-estimate the excited state population due to the failure to fully
recover spontaneous emission. For all regimes, Ehrenfest+R dynamics
can achieve detailed balance.\label{fig1}}
\end{figure}

Fig.~\ref{fig1}(a) shows that Ehrenfest dynamics can recover the
Boltzmann distribution for a large range of temperature. Surprisingly,
Ehrenfest dynamics results turn out to be more accurate when $\beta$
become larger (low temperature). This observation is the opposite
of Tully's analysis for electron\textendash nuclei equilibrium\citep{parandekar_mixed_2005}.
To rationalize this unexpected temperature dependence, we note that,
within the regime of electron\textendash nuclei interactions, one
can separate the system\textendash bath coupling strength from the
temperature of the bath. By contrast, for electron\textendash radiation
interactions, the coupling strength $A\left(\omega,\beta\right)$
itself depends on the temperature of the radiation fields and low
temperature generally leads to weak excitation. Now, if we recall
that Ehrenfest dynamics can almost recover spontaneous emission for
a weakly excited TLS\citep{li_mixed_2018,li_comparison_2019} (because
the overall effect of spontaneous emission is dominated by the effect
of self-interaction), we must conclude that, at low temperature, excitation
of the TLS by thermal radiation will be balanced by the Ehrenfest
decay rate (that is almost the correct spontaneous emission rate).

Next, as shown in Fig.~\ref{fig1}(b), at high temperature, the
equilibrium population obtained by Ehrenfest dynamics is larger than
the Boltzmann distribution. At high temperature, the TLS is strongly
excited by the incoming thermal radiation and the effects of vacuum
fluctuations become more important for spontaneous emission. And thus,
due to the failure of Ehrenfest dynamics to include vacuum fluctuations,
the Ehrenfest decay rate will be insufficient to balance with the
incoming thermal radiation. As a result, Ehrenfest dynamics predict
equilibrium populations that are too large and do not satisfy detailed
balance. 

Now, we turn our attention to the results of the coupled Maxwell\textendash Bloch
equations. Due to the double-counting of the self-interaction effect,
the coupled Maxwell\textendash Bloch equations almost always produce
a smaller equilibrium population than the Boltzmann distribution.
Following a similar analysis as for Ehrenfest dynamics, at high temperature
($\beta\hbar\omega_{0}<0.5$), the TLS will be strongly excited and
the effect of vacuum fluctuations should dominate the overall effect
of spontaneous emission. Therefore, the results of the coupled Maxwell\textendash Bloch
equations become closer to the Boltzmann distribution.

Finally, the equilibrium populations as attained by the Ehrenfest+R
approach agrees with the Boltzmann distribution for the whole range
of temperature. The success of the Ehrenfest+R approach demonstrates
that the ``+R'' correction provides a subtle patch for mean-field
Ehrenfest dynamics to achieve detailed balance by recovering spontaneous
emission. In contrast to the coupled Maxwell\textendash Bloch equation,
Ehrenfest+R dynamics incorporate the correct dissipation rate that
only account for vacuum fluctuations. Note that we require only a
quantum system and an ensemble of classical EM fields to fully satisfy
detailed balance, and we never have a quantized EM field or any other
QED calculation.

\section{Conclusion\label{sec5}}

In conclusion, we have shown that mixed quantum\textendash classical
electrodynamics can recover detailed balance when satisfying the requirements
of: (a) an appropriate classical model for thermal radiation fields
and (b) an accurate treatment of spontaneous emission. For (a), we
employ random electrodynamics to classically model thermal radiation
fields as boundary conditions. For (b), the Ehrenfest+R approach can
capture spontaneous emission quantitatively, including the self-interaction
and vacuum fluctuations. With this framework, unlike standard Ehrenfest
dynamics and the Maxwell\textendash Bloch equations, Ehrenfest+R dynamics
can achieve the correct Boltzmann distribution by balancing the optical
excitation of the incoming thermal radiation and the overall effect
of spontaneous emission. 

Given these promising results, we can envisage many exciting developments.
First, it will be interesting to evaluate the capability of the semiclassical
scheme to capture thermal equilibrium for a set of spatially separated
quantum emitters or a $N$-level quantum system\citep{parandekar_detailed_2006,tully_perspective:_2012}.
Such models can be employed for studying superradiant thermal emitter
assemblies\citep{zhou_analog_2015,mallawaarachchi_tuneable_2017}.
Second, with mixed quantum\textendash classical electrodynamics, one
will soon be able to simulate exciting phenomena in the field of nanoplasmonics,
including thermal excitation of surface plasmon \citep{liu_thermal_2014,guo_thermal_2014}
and thermalization of plasmon\textendash exciton polaritons\citep{rodriguez_thermalization_2013}.
This work should pave the way to many new applications of mixed quantum\textendash classical
electrodynamics.

\section*{Acknowledgment}

This material is based upon work supported by the U.S. Department
of Energy, Office of Science, Office of Basic Energy Sciences under
Award Number DE-SC0019397. The research of AN is supported by the
Israel-U.S. Binational Science Foundation.

\appendix

\section{Derivation of the discrete spectral density for thermal radiation\label{appendix_A}}

Within the theory of random electrodynamics\citep{boyer_random_1975,milonni_semiclassical_1976,boyer_understanding_2016,boyer_equilibrium_2018},
thermal radiation can be described as a sum over transverse plane
waves with a characteristic thermal spectrum corresponding to the
blackbbody radiation spectrum. We consider a large space of volume
$V$ with an incoming-wave boundary condition\citep{boyer_understanding_2016}.
The electric field can be written as a sum over all wave vectors $\mathbf{k}$
and polarizations $\lambda=1,2$:
\begin{equation}
\mathbf{E}_{\mathrm{in}}\left(\mathbf{r},t\right)=\sum_{\mathbf{k},\lambda}\boldsymbol{\epsilon}_{\mathbf{k},\lambda}\sqrt{\frac{2U\left(\omega,\beta\right)}{\epsilon_{0}V}}\cos[\mathbf{k}\cdot\mathbf{r}-\omega t+\theta_{\mathbf{k},\lambda}],\label{eq:Boyer_Ein}
\end{equation}
where $\theta_{\mathbf{k},\lambda}$ is a random phase chosen for
each $\mathbf{k},\lambda$. The magnetic field is obtained from $\frac{\partial}{\partial t}\mathbf{B}_{\mathrm{in}}\left(\mathbf{r},t\right)=-\boldsymbol{\nabla}\times\mathbf{E}_{\mathrm{in}}\left(\mathbf{r},t\right)$.
Here, $\boldsymbol{\epsilon}_{\mathbf{k},\lambda}$ denotes the polarization
unit vectors of the classical EM fields such that $\boldsymbol{\epsilon}_{\mathbf{k},1}\bot\boldsymbol{\epsilon}_{\mathbf{k},2}$
and $\mathbf{k}\bot\boldsymbol{\epsilon}_{\mathbf{k},\lambda}$ for
$\lambda=1,2$. As is common, we assume that the radiation fields
are isotropic. $U\left(\omega,\beta\right)$ is the energy per mode
of random electromagnetic radiation ($\omega=c\left|\mathbf{k}\right|=ck$).
Following Ref.~\onlinecite{boyer_understanding_2016}, we use the
Planck radiation spectrum given by 
\begin{equation}
U\left(\omega,\beta\right)=\frac{\hbar\omega}{e^{\beta\hbar\omega}-1}.\label{eq:Planck}
\end{equation}
Note that we do not include the zero-point energy term ($\frac{1}{2}\hbar\omega$)
in Eq.~\eqref{eq:Planck}. For random electrodynamics, as shown by
Boyer\citep{boyer_derivation_1969}, the zero-point energy can be
derived from the effects of the walls. In this paper, however, the
effect of zero-point energy is ignored in the boundary conditions,
but spontaneous emission is included heuristically in our semiclassical
treatments. 

From Eq.~\ref{eq:Boyer_Ein}, we can show that the statistical properties
of the random electrodynamic fields agree with QED when averaged over
the random phase $\theta$. First, the average electric field of the
incoming thermal radiation is zero, $\left\langle \mathbf{E}_{\mathrm{in}}\right\rangle _{\theta}=0$.
Second, the average intensity is associated with the energy density
\begin{equation}
\left\langle \left|\mathbf{E}_{\mathrm{in}}\right|^{2}\right\rangle _{\theta}=\sum_{\mathbf{k},\lambda}\frac{U\left(\omega,\beta\right)}{\epsilon_{0}V}.
\end{equation}
Here, we use the random phase correlation $\left\langle \cos\left[\mathbf{k}\cdot\mathbf{r}-\omega t+\theta_{\mathbf{k},\lambda}\right]\cos\left[\mathbf{k}^{\prime}\cdot\mathbf{r}^{\prime}-\omega^{\prime}t+\theta_{\mathbf{k}^{\prime},\lambda^{\prime}}\right]\right\rangle _{\theta}=\frac{1}{2}\delta^{3}\left(\mathbf{k}-\mathbf{k}^{\prime}\right)\delta_{\lambda\lambda^{\prime}}$.
For a very large $V$ and isotropic fields, we can replace the summation
using 
\begin{equation}
\sum_{\mathbf{k},\lambda}\rightarrow2\frac{V}{\left(2\pi\right)^{3}}\int k^{2}\mathrm{d}k\mathrm{d}\Omega=\sum_{j=1}^{{\cal N}}\frac{V}{\pi^{2}c^{3}}\omega_{j}^{2}d\omega,
\end{equation}
where we discretize the integral by $\omega_{j}=jd\omega$ with small
enough $d\omega$. Thus, the average intensity can be written as a
summation of the energy density within the frequency range $\left(\omega_{j},\omega_{j}+d\omega\right)$
\begin{equation}
\left\langle \left|\mathbf{E}_{\mathrm{in}}\right|^{2}\right\rangle _{\theta}=\sum_{j}\frac{\omega_{j}^{2}}{\pi^{2}\epsilon_{0}c^{3}}U\left(\omega_{j},\beta\right)d\omega.\label{eq:intensity}
\end{equation}
Note that both $\left\langle \mathbf{E}_{\mathrm{in}}\right\rangle _{\theta}$
and $\langle\left|\mathbf{E}_{\mathrm{in}}\right|^{2}\rangle_{\theta}$
are independent of position and time for thermal radiation fields. 

For a TLS emitter at $\mathbf{r}=0$ and $\mathbf{d}\left(\mathbf{r}\right)\rightarrow\mu\delta^{3}\left(r\right)\boldsymbol{\epsilon}_{z}$,
we can write the electric coupling as 
\begin{equation}
\mu\boldsymbol{\epsilon}_{z}\cdot\mathbf{E}_{\mathrm{in}}\left(0,t\right)=\mu\sum_{\omega}A\left(\omega,\beta\right)\cos\left[\omega t+\theta_{\omega}\right]
\end{equation}
To find the coupling strength $A\left(\omega,\beta\right)$, we utilize
the isotropic property of the EM field, such that the $z$-component
of the random electric field should satisfy 
\begin{equation}
\left\langle \left|\boldsymbol{\epsilon}_{z}\cdot\mathbf{E}_{\mathrm{in}}\right|^{2}\right\rangle _{\theta}=\frac{1}{3}\left\langle \left|\mathbf{E}_{\mathrm{in}}\right|^{2}\right\rangle _{\theta}.
\end{equation}
Next, we use the random phase correlation $\left\langle \cos\left[\omega t+\theta_{\omega}\right]\cos\left[\omega^{\prime}t+\theta_{\omega^{\prime}}\right]\right\rangle _{\theta}=\frac{1}{2}\delta\left(\omega-\omega^{\prime}\right)$
to express
\begin{equation}
\left\langle \left|\boldsymbol{\epsilon}_{z}\cdot\mathbf{E}_{\mathrm{in}}\right|^{2}\right\rangle _{\theta}=\sum_{\omega}\frac{1}{2}\left|A\left(\omega,\beta\right)\right|^{2}.
\end{equation}
Therefore, from Eq.~\eqref{eq:Planck} and \eqref{eq:intensity},
we can conclude 
\begin{equation}
\left|A\left(\omega,\beta\right)\right|^{2}=\frac{2\hbar}{3\pi^{2}\epsilon_{0}c^{3}}\frac{\omega^{3}}{e^{\beta\hbar\omega}-1}d\omega.
\end{equation}

\bibliographystyle{aipnum4-1}
\bibliography{MyLibrary}

\end{document}